\documentclass[preprint,showpacs,preprintnumbers,amsmath,amssymb, superscriptaddress,longbibliography,nofootinbib]{revtex4-1}

\usepackage{textcomp}
\usepackage{makeidx}
\usepackage{amsmath}
\usepackage{subfig}
\usepackage{amssymb}
\usepackage{hyperref}
\usepackage{graphicx}
\usepackage{epstopdf}
\usepackage{color}
\usepackage{soul}

\begin{document}

\title{Quantized $p$-form Gauge Field in D-dimensional de Sitter Spacetime}

\author{E.W.D. Dantas \footnote{E-mail: emanuelwendelldd@gmail.com}}\affiliation{Departamento de F\'isica, Universidade Federal do Cear\'a, Caixa Postal 6030, Campus do Pici, 60455-760 Fortaleza, Cear\'a, Brazil}

\author{G. Alencar \footnote{E-mail: geova@fisica.ufc.br}}\affiliation{Departamento de F\'isica, Universidade Federal do Cear\'a, Caixa Postal 6030, Campus do Pici, 60455-760 Fortaleza, Cear\'a, Brazil}

\author{I. Guedes \footnote{E-mail: geova@fisica.ufc.br}}\affiliation{Departamento de F\'isica, Universidade Federal do Cear\'a, Caixa Postal 6030, Campus do Pici, 60455-760 Fortaleza, Cear\'a, Brazil}

\author{Milko Estrada \footnote{E-mail: milko.estrada@gmail.com}}\affiliation{Facultad de Ingeniería y Empresa, Universidad Católica Silva Henríquez, Chile}

\date{\today}

\begin{abstract}
In this work, we utilize the dynamic invariant method to obtain a solution for the time-dependent Schrödinger equation, aiming to explore the quantum theory of a $p$-form gauge field propagating in $D$-dimensional de Sitter spacetimes.
Thus, we present a generalization, through the use of $p$-form gauge fields, of the quantization procedure for the scalar, electromagnetic, and Kalb-Ramond fields, all of which have been previously studied in the literature. We present an exact solution for the $p$-form gauge field when $D=2(p+1)$, and we highlight the connection of the $p=4$ case with the chiral $N=2$, $D=10$ superstring model. We could observe particle production for $D \neq 2(p+1)$ because the solutions are time-dependent. Additionally, observers in an accelerated co-moving reference frame will also experience a thermal bath. This could have significance in the realm of extra-dimensional physics and presents the intriguing prospect that precise observations of the Cosmic Microwave Background might confirm the presence of additional dimensions.
\end{abstract}

\maketitle

\section{Introduction}
The Quantum Field Theory in flat (Minkowski) spacetimes is one of the most successful theories in physics. It serves as the foundation upon which the Standard Model of particles is constructed and provides a quantum description of the strong, weak, and electromagnetic forces. On the other hand, gravity is described by Einstein's General Theory of Relativity, a classical theory, which has also proven to be very successful \cite{Will:2014kxa}. However, it is well understood that General Relativity remains an incomplete theory. Attempts to incorporate gravity into the Standard Model have proven to be non-renormalizable, as it requires an infinite number of parameters to do so.

Quantization of gravity is one of the most difficult and arduous challenges in modern physics and mathematics. Various approaches to quantizing gravity have been developed, with the most well-known being String Theory, which achieves quantization along with unification with the other three forces.  However, experimental validation of the theory poses a significant obstacle due to the extremely weak quantum effects of gravity. This fact allows for creativity in the search for observable characteristics of the theory \cite{Hossenfelder:2009nu}

Despite the success of quantum field theory in flat spaces, and some theoretical achievements of string theory, there are still several problems related to the behavior of fields in curved spaces in a quantum theory at cosmological scales. For example, interesting results have been achieved in a background of time-dependent fields such as for example black Hawking radiation \cite{Hawking:1974rv}, which predicts the evaporation of black holes. Closely related is the Unruh effect \cite{Padmanabhan:2003gd}, which suggests that an accelerated particle would perceive a thermal bath, and the dynamical Casimir effect \cite{Leonhardt:2023szx}, which predicts particle creation by an accelerated mirror. Another notable outcome is particle creation by the vacuum, which has been extensively researched \cite{Haouat:2011uy,Grib:1974ym,Basler:1988ra,Pavlov:2011sgb,Parker:1968mv,Parker:1969au,Parker:1971pt}, with the confirmation of the Schwinger effect expected soon \cite{Berdyugin_2022}.

On the other hand, the investigation of de Sitter spacetime gains significance when contemplating a $\Lambda$CDM model of the cosmos. In the current epoch, the universe can be roughly characterized by de Sitter spacetime, wherein matter decays with volume while the cosmological constant remains constant. For $t \gg H^{-1}$, the universe tends to be effectively described by the de Sitter model.
In this connection, the study of quantum effects of a massive scalar field in de Sitter spacetime was examined in Ref. \cite{Lopes}, where the authors utilized exact linear invariants and the Lewis and Riesenfeld method \cite{Lewis:1968tm} to derive the corresponding Schrödinger states based on solutions of a second-order ordinary differential equation. Additionally, they formulated Gaussian wave packet states and computed the quantum dispersion and correlations for each mode of the quantized scalar field. 

The quantization of certain types of fields has been explored in the literature. For the scalar field \cite{Alencar:2011an}, it has been shown that the conformability of the system is tied to the choice of the curvature parameter. Similarly, the electromagnetic field \cite{Alencar:2011nw} proves to be conformal in $D=4$, as expected. Additionally, it has been observed that the Kalb-Ramond field is conformal in $D=6$ \cite{Alencar:2012yx}.

On the other hand, various branches of theoretical physics have provided indications of the possible existence of extra dimensions. Examples include string theory, higher-dimensional black holes, or brane world models. It is worth mentioning that higher-dimensional FRW scenarios, including the de Sitter scenario, and their particle creation, have also garnered attention in recent years \cite{Aygun:2016hre,Alfedeel:2023aky}. In this context, as indicates reference \cite{Weinberg:1995mt}:  p-form gauge fields play a significant role in theories involving extra dimensions. For example, within string theories in 26 dimensions, a low-energy normal mode of the string is represented by a two-form gauge field $A_{\mu \nu}$. However, in four-dimensional spacetimes, p-forms do not introduce new possibilities.

In this study, we will employ the method developed  by Lewis and Riesenfeld \cite{Lewis:1968tm} to find a solution for the equations of motion governing a $p$-form gauge field in a $D$-dimensional de Sitter spacetime. Subsequently, we will proceed with its quantization by determining a solution to the 'auxiliary' equation \cite{Milne,Pinney}. Additionally, we will discuss the potential physical implications of our analysis for various extra-dimensional scenarios of interest in physics.

\section{Equations of Motion and its Decomposition}

We will employ the Friedmann-Lema\^{i}tre-Robertson-Walker(FLRW, for short) metric in $D$ dimensions($1 + (D-1)$), which has the line element $ds^{2} = - dt^{2} + a^{2}(t) d \mathbf{x} \cdot d \mathbf{x}$.
The action is given by
\begin{equation}
S = \frac{1}{2(p+1)!} \int d^{D}x \sqrt{-g} F^{\mu\mu_1 \cdots \mu_p} F_{\mu\mu_1 \cdots \mu_p},
\end{equation}
where the field strength is $F_{\mu \mu_1 \cdots \mu_p}=\partial_ {[\mu} A_{\mu_1 \cdots \mu_p]}$, and $A_{\mu_1 \cdots \mu_p}$ is the $p$-form gauge field. 

Although there are various paths to perform the decomposition of the field in its normal modes, as its seen in \cite{Alencar:2011an} for example, there are no drawbacks to performing it directly from the equations of motion.
\begin{equation}\label{EMFp}
\partial_{\nu}(\sqrt{-g}g^{\mu\nu}g^{\mu_1\nu_1} \cdots g^{\mu_p\nu_p}F_{\mu\mu_1 \cdots \mu_p})=0.
\end{equation}

To simplify this equation we can fix the gauge by means of the transverse condition $\partial^{i_1}A_{i_1 \cdots i_p}=0$ and the invariance of the gauge allows us to fix $A_{0i_2 \cdots i_p}=0$. Putting these two conditions into (\ref{EMFp}) will leads us to the following equation
\begin{equation}\label{EMgauge}
\ddot{A}_{i_1 \cdots i_p} + (D-2p-1) \frac{\dot{a}}{a} \dot{A}_{i_1 \cdots i_p} - \frac{1}{a^{2}} \nabla^{2} A_{i_1 \cdots i_p} = 0.
\end{equation}

Now we take the standard approach by tackling this equation with the usual normal modes decomposition
\begin{equation}\label{decomp}
A_{i_1 \cdots i_p} (\mathbf{x},t) = \sum^{\alpha}_{\epsilon =1} \int \frac{d^{D-1}k}{(2\pi)^{D-1}} f_{i_1 \cdots i_p}^{\epsilon} \left( r_{1}^{\epsilon}(t) e^{i\mathbf{k}\cdot \mathbf{x}} + r_{2}^{\epsilon}(t) e^{-i\mathbf{k}\cdot \mathbf{x}} \right),
\end{equation}
with $f^{\epsilon}_{i_1 \cdots i_p}$ representing the various polarizations obeying the gauge
condition $k^{i_1} f^{\epsilon}_{i_1 \cdots i_p}=0$ and $\displaystyle \alpha = \frac{(D-2)(D-3)}{2}$. Substituting (\ref{decomp}) in (\ref{EMgauge}) we finally arrive in our desired equation for the modes
\begin{equation}\label{EMmodes}
\ddot{r} + (D-2p-1) \frac{\dot{a}}{a} \dot{r} + \frac{k^{2}}{a^{2}} r = 0.
\end{equation}

Here, we have omitted all indices attached to $r$, as equation (\ref{EMmodes}) remains the same for all of them. 

Now, in order to find a solution to this equation we will examine the Hamiltonian for the harmonic oscillator with a time-dependent mass and frequency, $m = m(t)$ and $\omega = \omega(t)$, respectively,
\begin{equation}\label{Hamiltoniano}
H(t) = \frac{p^{2}}{2m} + \frac{m \omega^{2}}{2} q^{2}.
\end{equation}
where $p$ and $q$ are now considered operators and obey the relation $[q,p]=i \hbar$. The equations of motion are trivially obtained and are given by
\begin{equation} \label{EcuacionDeMovimiento}
\ddot{q} + \frac{\dot{m}}{m} \dot{q} + \omega^{2} q = 0,
\end{equation}
which has a striking resemblance to (\ref{EMmodes}). In view of this similarity we can consider our system as being a time-dependent harmonic oscillator with mass given by $m=a^{D-2p-1}$ and frequency $\displaystyle \omega=\frac{k}{a}$.

\section{Quantization of the $p$-form gauge field in the de Sitter Space-time}
In this section we will discuss the tool to obtain a solution to the time-dependent harmonic oscillator.

We search for an $I(t)$ with the requirement of it being invariant of the Hamiltonian of equation \eqref{Hamiltoniano} \cite{Carinena}
\begin{equation}\label{inv}
\frac{dI}{dt} = \frac{\partial I}{\partial t} + \frac{1}{i \hbar} \left[ I, H \right]=0,
\end{equation}
with real eigenvalues, which makes it Hermitian. It turns out that the solution $|\psi_{n}\rangle$ for the Schr\"odinger equation by means of this invariant (\ref{inv}) cannot be completely determined just by the assumption that it needs to be Hermitian. We need to specify the phase $\theta$ of the invariant's eigenstates $ \left| n, t \right. \rangle$, assumed to form a complete orthonormal basis for $I$. Then the exact solution is given by
\begin{equation}\label{schrstate}
\left| \psi_{n}\rangle \right. = e^{i \theta_{n}(t)} \left| n, t \rangle \right.,
\end{equation}
where the phase $\theta_{n}(t)$ needs to satisfy the following equation \cite{Lewis:1968tm}
\begin{equation} \label{CondicionDeFase}
\hbar \frac{d\theta_{n}(t)}{dt} = \left \langle n, t \left| \left( i \hbar \frac{\partial}{\partial t} - H(t) \right) \right| n, t \right \rangle.
\end{equation}

Since the invariant is of our choice(given that we follow the conditions set upon it), we consider the following choice \cite{Carinena}
\begin{equation}\label{invchoice}
I = \frac{1}{2} \left[ \left( \frac{q}{\rho} \right)^{2} + (\rho p - m \dot{\rho}q)^{2} \right],
\end{equation}
where $q(t)$ satisfies equation \eqref{EcuacionDeMovimiento} and where $\rho = \rho(t)$ satisfies the generalised Milne-Emarkov-Pinney (MP) \cite{Milne,Pinney} equation 
\begin{equation}\label{auxiliary}
\ddot{\rho} + \frac{\dot{m}}{m} \dot{\rho} + \omega^{2} \rho = \frac{1}{m^{2} \rho^{3}},
\end{equation}
where the choice for (\ref{invchoice}) comes from assuming it to be of a quadratic form in $q$ and $p$, so it would closely resemble (\ref{Hamiltoniano}).

To perform the quantisation we take the standard route of considering the (time-dependent) creation $b^{\dagger}(t)$ and annihilation $b(t)$ operators defined as
\begin{equation}
b^{\dagger} = \sqrt{ \frac{1}{2 \hbar} } \left[ \left( \frac{q}{\rho} - i \left( \rho p - m \dot{\rho} q \right) \right)\right];
\end{equation}
\begin{equation}
b = \sqrt{ \frac{1}{2 \hbar} } \left[ \left( \frac{q}{\rho} + i \left( \rho p - m \dot{\rho} q \right) \right) \right],
\end{equation}
constructed so that $\left[ b, b^{\dagger} \right] = 1$, with the usual properties
$$
b \left| n, t \right. \rangle = \sqrt{n} \left| n-1, t \right. \rangle
$$
$$
b^{\dagger} \left| n, t \right. \rangle = \sqrt{n+1} \left| n+1, t \right. \rangle
$$
$$
b^{\dagger} b \left| n, t \right. \rangle = n \left| n, t \right. \rangle , 
$$
and we are assuming the eigenvalues for $I$ to be discrete. This allows us to write the following eigenvalue equation for the equation (\ref{invchoice})
\begin{equation}
    I \left| n, t \right. \rangle = \lambda_n \left| n, t \right. \rangle
\end{equation}
and we see that the eingenvalues of $I$ are given by
\begin{equation}
 \lambda_n = \hbar \displaystyle \left( n + \frac{1}{2} \right)   
\end{equation}
where $n=b^{\dagger} b$. These assumptions follow the same path we make to quantise the Hamiltonian, so its safe to assume that the eigenstates for the invariant $I(t)$ are indeed related to the Hamiltonian's by means of (\ref{schrstate}).

The Schrödinger equation is given by:
\begin{equation}
    i\frac{\partial \psi(q, t)}{\partial t} = H(t) \psi(q, t)
\end{equation}
Lewis and Riesenfeld showed that the the general solution of the  Schrödinger equation is given by:
\begin{equation}
\displaystyle    \psi(q, t) = \sum_{n} c_{n} \left| \psi_{n}\rangle \right.
\end{equation}
where $c_n$ are time-independent coefficients and where $\left| \psi_{n}\rangle \right.$ satisfies equation \eqref{schrstate} and phase $\theta_{n}(t)$ satisfy the equation \eqref{CondicionDeFase}.

By using a unitary transformation and following the steps outlined in reference \cite{Carinena}, the normalised solution for the time-dependent harmonic oscillator is then written as
\begin{equation}\label{wavefunction}
\left| \psi_{n}\rangle \right. (q, t) = e^{i \theta_{n}} \left( \frac{1}{ (\pi \hbar)^{1/2} n! 2^{n} \rho} \right)^{1/2} \exp \left \lbrace \frac{im}{2 \hbar} \left[ \frac{\dot{\rho}}{2 \hbar} + \frac{i}{m \rho^{2}}\right] q^{2} \right \rbrace H_{n} \left( \frac{q}{\rho \sqrt{\hbar}} \right)
\end{equation}
where $H_{n}$ are the Hermite polynomials of order $n$, and the phase $\theta_{n}(t)$ from (\ref{CondicionDeFase}) now reads
\begin{equation} \label{fasetheta}
    \theta_{n}(t) = -\left(n + \frac{1}{2}\right) \int_{t_{0}}^{t} \frac{1}{m(t') \rho^{2}}  \, dt' .
\end{equation} 
Hence, quantizing the time-dependent harmonic oscillator hinges on identifying a solution to the corresponding MP equation \eqref{auxiliary}, which will be incorporated into equation \eqref{wavefunction}. It is worth mentioning that a solution to this nonlinear equation consists of a nonlinear combination of solutions to the linear case \cite{Prince}. Notice that the linear form of the MP equation mirrors our classical equation of motion \eqref{EcuacionDeMovimiento}. Hence, discovering solutions to \eqref{EcuacionDeMovimiento} enables us to uncover the sought-after solution to the problem.

Let's now apply the process of quantizing the $p$-form gauge field. With $m = a^{D-2p-1}$ 
and $\omega(t) = \displaystyle \frac{k}{a}$ the auxiliary Milne-Emarkov-Pinney equation is
\begin{equation}\label{MEP}
\ddot{\rho} + (D-2p-1) \frac{\dot{a}}{a} \dot{\rho} + \frac{k^2}{a^2} \rho = \frac{1}{a^{2(D-2p-1)} \rho^{3}}.
\end{equation}
As mentioned earlier, in order to obtain the solution for equation \eqref{MEP}, we will initially seek solutions to the classical equation \eqref{EMmodes}. Now, by means of a change to the conformal time $\eta$ by setting $dt=a d \eta$ and $r= \Omega\bar{r}$, we get from (\ref{EMmodes})
\begin{equation}\label{generaleq}
 \bar{r}'' + \left( 2a \frac{\dot{\Omega}}{\Omega} - \dot{a} + (D-2p-1) \dot{a} \right) \bar{r}' + \left( k^{2} + a^{2} \frac{\ddot{\Omega}}{\Omega} + (D-2p-1)a \dot{a} \frac{\dot{\Omega}}{\Omega} \right)\bar{r} = 0,
\end{equation}
where prime (') and dot (.) represent differentiation with respect to the conformal time $\eta$ and $t$, respectively.If we make the choice $\Omega = a^{-(D-2p-1)/2}$ we get 
\begin{equation}
 \bar{r}'' - \dot{a} \bar{r}' + \left[ k^{2} - \frac{(D-2p-1)}{2} a \ddot{a} + \frac{(D-2p-1)(D-2p-3)}{4} \dot{a}^{2} \right] \bar{r} = 0,
\end{equation}

We'd like to emphasize that when $D=2p+1$, certain terms cancel out, leading to a simplified equation contingent upon the choice of parameter $a$. Let us consider, the de Sitter spacetime where $a = e^{Ht}$. The expressions for $a$, $\dot{a}$ and $\ddot{a}$ are reduced to
\begin{equation}
a = -\frac{1}{H \eta}, \; \; \; \; \dot{a} = -\frac{1}{\eta}, \; \; \; \; \ddot{a} = -\frac{H}{\eta},
\end{equation}
and we finally obtain
\begin{equation}
\frac{d^{2} \bar{r}}{d(k \eta)^{2}} + \frac{1}{(k \eta)} \frac{d\bar{r}}{d(k \eta)} + \left( 1 - \frac{\nu^{2}}{(k \eta)^{2}} \right)\bar{r} = 0,
\end{equation}
where $\nu = \displaystyle \frac{D-2p-1}{2}$. This equation is Bessel's equation which has two linearly independent solutions, given by $J_{\nu} \left( k \left| \eta \right| \right)$ and $Y_{\nu} \left( k \left| \eta \right| \right)$, the Bessel's functions of first and second kind, respectively. Now employing our earlier redefinition of $r= \Omega\bar{r}$, with $\Omega = a^{-(D-2p-1)/2}$, we obtain that two linearly independent solutions for $r$ are:

\begin{equation}
    r = \begin{cases} 
&a^{-(D-2p-1)/2} J_{\nu} \left( k \left| \eta \right| \right) \\ 
&a^{-(D-2p-1)/2}  Y_{\nu} \left( k \left| \eta \right| \right) 
\end{cases}
\end{equation}
Following references \cite{Bertoni:1997qb,Finelli:1999dk}, a particular solution of the equation \eqref{MEP} is:
\begin{equation}
\rho = - (H \eta)^{(D-2p-1)/2} \left[ A J_{\nu}^{2} + B Y_{\nu}^{2} + 2 \left( AB - \frac{\pi^{2}}{4H^{2}} \right)^{1/2 }J_{\nu} Y_{\nu} \right]^{1/2} ,
\end{equation}
where $A$ and $B$ are real constants. The determination of these constants is intricately tied to our vacuum selection. This arises from the non-uniqueness in constructing particle states and selecting the vacuum in curved spaces like the one employed in this scenario. This holds significance because the generation of particles can only be deduced once we have selected a vacuum for comparison with our physical solution. 

In our scenario, a suitable choice corresponds to the Bunch-Davies vacuum, which aligns with the adiabatic vacuum for very early times ($t \rightarrow -\infty$) or the adiabatic vacuum for wavelengths much smaller than the de Sitter horizon $H^{-1}$. With these assumptions, the values of the constants are $A=B=\pi/2H$ \cite{Bertoni:1997qb}, and $\rho$ is given by
\begin{equation}\label{solution}
\rho = \left( H \left| \eta \right| \right)^{(D-2p-1)/2} \sqrt{\frac{\pi}{2H}}\left( J_{\nu}^{2} + Y_{\nu}^{2}\right)^{1/2}.
\end{equation}

Now that we have finally found the general solution to the Milne-Emarkov-Pinney equation (\ref{MEP}) in a de Sitter scenario, we can substitute it in the expression for the solution of the harmonic oscillator with mass and frequency dependent on time (\ref{wavefunction}). This concludes the quantisation of the $p$-form gauge field in a $D$-dimensional de Sitter background.

\section{Concluding Remarks}
In this work, we have presented a generalization of the quantization procedure, through the use of $p$-form gauge fields, for the scalar, electromagnetic, and Kalb-Ramond fields, all of which have been previously studied and referenced \cite{Alencar:2011an,Alencar:2011nw,Alencar:2012yx}. In this connection, we have obtained a solution to the Schrödinger equation using the method developed by Lewis and Reisenfeld \cite{Lewis:1968tm}, applied to the quantization of the $p$-form field in a $D$-dimensional de Sitter space-time. A general solution for equation (\ref{generaleq}) is found to depend on the scale factor present in the FLRW space-time and was obtained in the particular case of de Sitter space-time, which is significant because our Universe today can be approximated as such, and in the far future, it would fully become one.

We can check that the equation (\ref{solution}) is constant for $D=2(p+1)$. This is in agreement with the previous works for $D=4$ and $p=1$ \cite{Alencar:2011nw,Finelli:1999dk}. Thus, for $D=2(p+1)$, for a (massless) photon, the initial adiabatic vacuum persists indefinitely, resulting in zero photon production within de Sitter spacetime, while its energy undergoes the redshift characteristic of radiation.

An interesting case where $D=2(p+1)$ arises for $p=4$ in a $10$-dimensional space-time, where $A_{\mu_1 \cdots \mu_p}$ corresponds to a $4$-form gauge field, while the field strength corresponds to a $5$-form. In this scenario, there will be no production of particles, and a co-moving accelerated observer will not experience a thermal bath. This specific dimension value indicates that the $4$-form exhibits conformal invariance, allowing for the straightforward solution of the time-dependent harmonic oscillator \eqref{wavefunction} in a de Sitter spacetime.
Since our field strength is going to be a $5$-form it will be dual to itself $F_{\mu_1 \cdots \mu_5} = *F_{\mu_1 \cdots \mu_5}$. The case where $D=10$ has garnered attention as it represents the critical dimension in superstring theory, and the $5$-form field strength naturally appears as a first-order approximation for the gravitational coupling constant of chiral $N=2$ $D=10$ supergravity \cite{Schwarz:1983wa}.

For $D \neq 2(p+1)$, it becomes evident that we can no longer have a constant solution for (\ref{solution}), regardless of our chosen cosmological model. In a future work, a more general solution could be computed for the case of de Sitter space-time, which undoubtedly has more interesting consequences. Specifically, we can observe particle production for $D \neq 2(p+1)$ because the solutions are time-dependent. Additionally, observers in an accelerated co-moving reference frame will also experience a thermal bath.

As mentioned in the introduction, gauge $p$-form fields in certain extra-dimensional scenarios exhibit physical properties that are not visible in four dimensions. Thus, an intriguing implication arises in the realm of extra-dimensional physics, which has garnered significant attention in certain scenarios involving de Sitter spacetimes. For instance, in higher-dimensional FRW scenarios and their associated particle creation \cite{Aygun:2016hre,Alfedeel:2023aky}, or in de-Sitter braneworld models \cite{Das:2007qn,Aros:2012xi}. Braneworld models where FRW branes possess a temperature have been investigated in reference \cite{Aros:2016wpv}. In braneworld models, our universe is conceptualized as a brane existing within a five-dimensional space. Consequently, in such a setup, a de Sitter spacetime would lead to particle production and the presence of a thermal bath for observers moving within this expanded space. Consequently, this could potentially contribute to an effective temperature within the membrane, suggesting the intriguing possibility that precise measurements of the Cosmic Microwave Background could reveal the presence of extra dimensions.

\section*{Acknowledgements}
 Milko Estrada is funded by the FONDECYT Iniciaci\'on 2023 folio 11230247. E.W.D. Dantas, I. Guedes and G. Alencar thank the Coordena\c{c}\~{a}o de Aperfei\c{c}oamento de Pessoal de N\'{i}vel Superior (CAPES).

\bibliography{mybib} 

\end{document}